\documentclass{moriond}

\usepackage[english]{babel}
\usepackage{amsmath,amssymb}
\usepackage{cleveref}
\usepackage{graphicx}
\usepackage[dvipsnames]{xcolor}
\usepackage[sort&compress]{natbib}
\usepackage{siunitx}

\bibpunct{[}{]}{,}{n}{}{,}

\def\be{\begin{equation}}
\def\ee{\end{equation}}
\def\bea{\begin{eqnarray}}
\def\eea{\end{eqnarray}}

\newcommand{\Photo}{\vspace{3mm}\includegraphics[height=35mm]{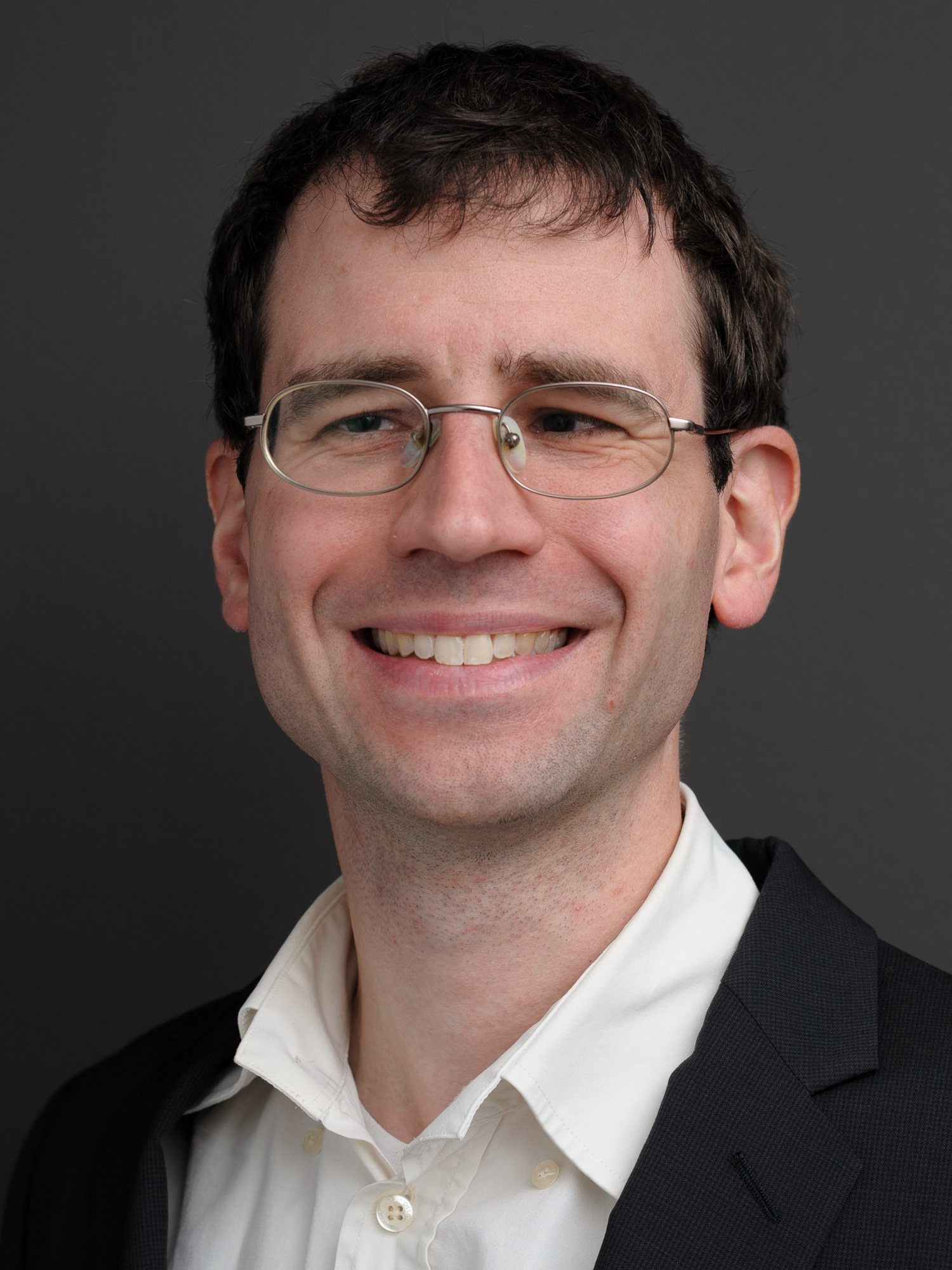}}
\newcommand{\iso}[2]{{\ensuremath{{}^{#2}}\ensuremath{\rm #1}}}

\begin{document}
\vspace*{4cm}
\title{THE NEUTRINO MAGNETIC MOMENT PORTAL}

\author{V.\ BRDAR$^{a,b}$,
        A.\ GRELJO$^{c,d}$,
        {\bf J.\ KOPP}$^{c,e}$ (speaker),
        T.\ OPFERKUCH$^c$}

\address{$^a$ Fermi National Accelerator Laboratory, Batavia, IL, 60510, USA \\
         $^b$ Northwestern University, Department of Physics \& Astronomy,
              Evanston, IL 60208, USA \\
         $^c$ Theoretical Physics Department, CERN, 1211 Geneva, Switzerland \\
         $^d$ Albert Einstein Center for Fundamental Physics,
              University of Bern, 3012 Bern, Switzerland \\
         $^e$ PRISMA+ Cluster of Excellence,
              Johannes Gutenberg-Universit\"{a}t Mainz, 55099 Mainz, Germany}

\maketitle\abstracts{
We discuss neutrino magnetic moments as a way of constraining physics beyond
the Standard Model. In fact, new physics at the TeV scale can easily generate
observable neutrino magnetic moments, and there exists a multitude of ways
of probing them. We highlight in particular direct dark matter detection
experiments (which are sensitive to neutrino magnetic moments because of the
predicted modifications to the solar neutrino scattering rate), stellar cooling,
and cosmological constraints.}

\section{Neutrino Magnetic Moments as a Probe for ``New Physics''}
\label{sec:mm}

With the particle physics world abuzz about the intriguing anomaly observed
in the magnetic moment of the muon, it is often forgotten that also the
muon's little siblings -- the neutrinos -- have magnetic moments. While
unobservably tiny in Standard Model, these magnetic moments may be significantly
enhanced in the presence of physics beyond the Standard Model, making them
excellent indirect probes of ``new physics''.  In these proceedings,
based largely on ref.~\cite{Brdar:2020quo}, we discuss neutrino
magnetic moments in the context of terrestrial, astrophysical, and cosmological
probes, and we comment on intriguing model building aspects.

\begin{figure}
  \centering
  \begin{tabular}{c@{\quad\quad}c}
    \includegraphics[width=0.3\textwidth]{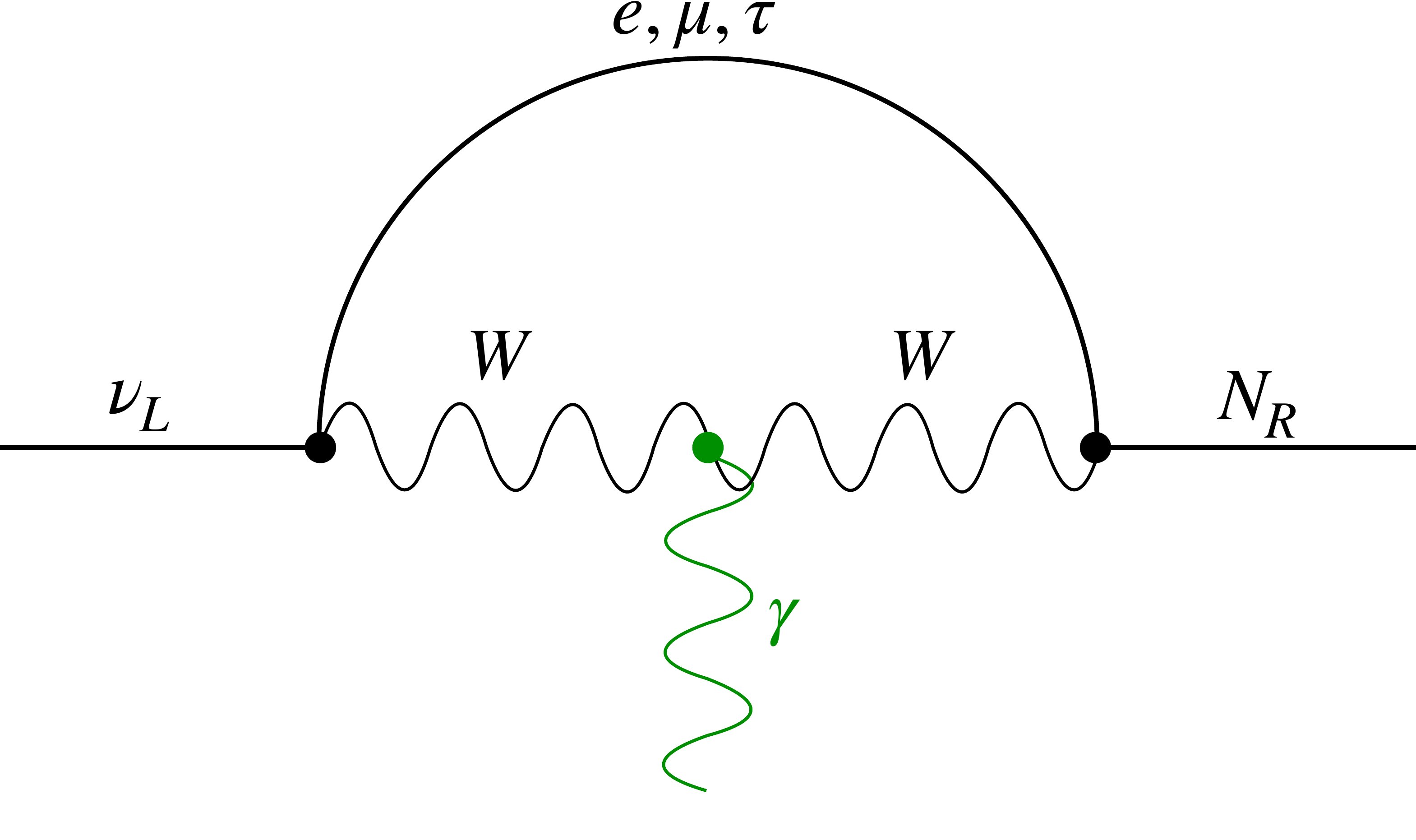} &
    \includegraphics[width=0.3\textwidth]{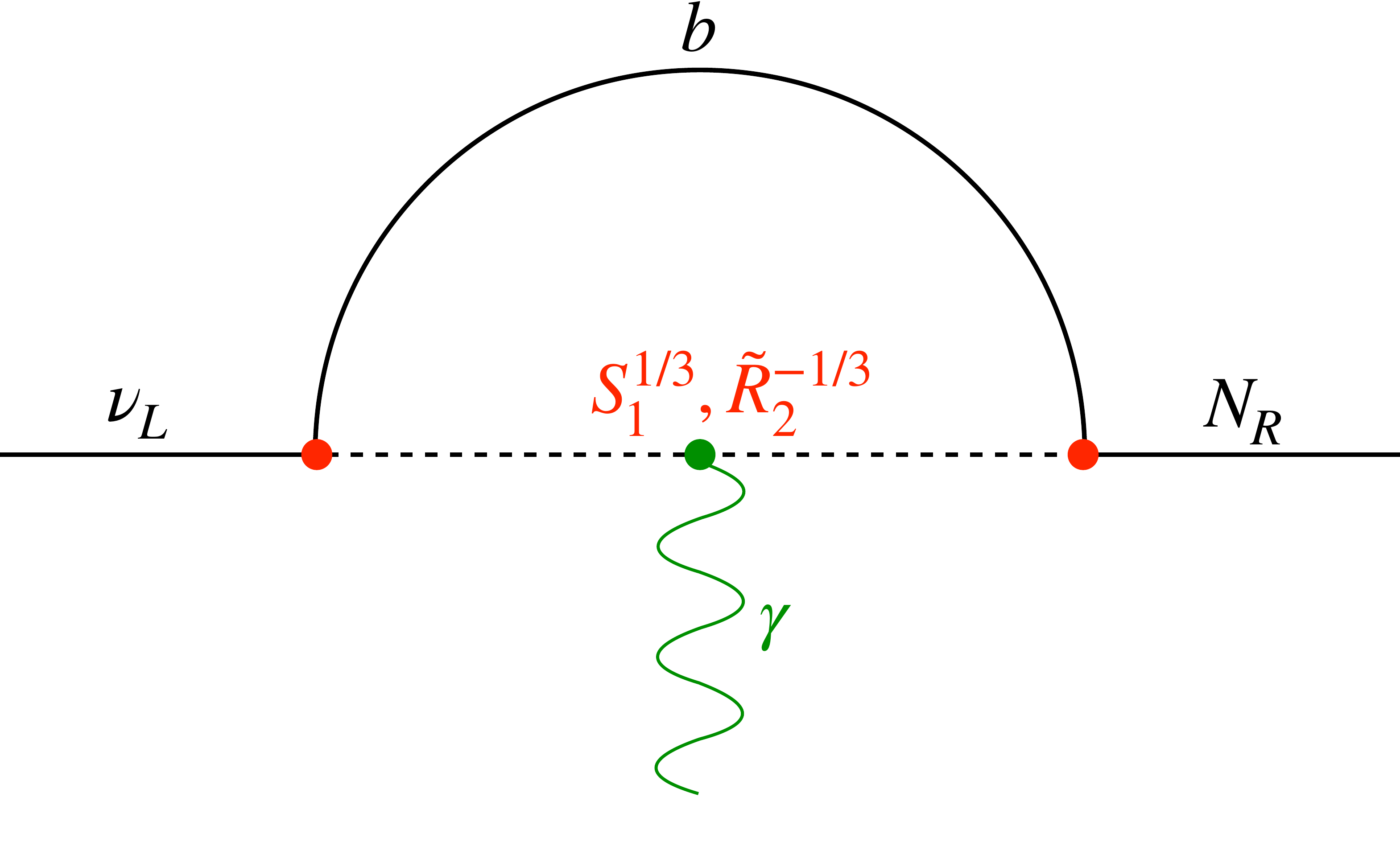} \\
    (a) & (b)
  \end{tabular}
  \caption{Diagrams contributing to the neutrino magnetic moment (a)
    in the SM and (b) in a leptoquark model.}
  \label{fig:feynman}
\end{figure}

At the effective field theory (EFT) level, a neutrino magnetic moment
is described by an operator of the form
\begin{align}
  \mathcal{L} \supset \frac{1}{2} \mu_\nu^{\alpha\beta} \,
                      \bar\nu_L^\alpha \sigma^{\mu\nu} N_R^\beta F_{\mu\nu} \,,
  \label{eq:mm}
\end{align}
where $\mu_\nu^{\alpha\beta}$ is the magnetic moment (a constant of mass dimension
$-1$), $\nu_L^\alpha$ and $N_R^\beta$ are left-handed and right-handed
neutrino fields, respectively, $F^{\mu\nu}$ is the electromagnetic field
strength tensor, and $\alpha$, $\beta$ are flavor indices. The operator in
\cref{eq:mm} is typically generated via loop diagrams such as the ones shown in
\cref{fig:feynman} as an example.  Note in particular that because magnetic moment
operators couple left-handed states to right-handed ones, they are sensitive to
the mechanism of neutrino mass generation. For Dirac neutrinos, the $N_R$ are
new states that are as light as the $\nu_L$. For Majorana neutrinos, the
anti-particles of the $\nu_L$ can play the role of the $N_R$; in that case,
the flavor-diagonal components of $\mu_\nu^{\alpha\beta}$ vanish and only
the off-diagonal ones (``transition magnetic
moments'') are non-zero as can be verified with a bit of Dirac algebra.  In
both cases, the coupling of the right-handed states to the $W$ -- which is
needed for the generation of $\mu_\nu^{\alpha\beta}$ in the SM, see
\cref{fig:feynman} (a) -- is suppressed by their mass, explaining the smallness
of $\mu_\nu^{\alpha\beta}$ in the SM \cite{Fujikawa:1980yx, Lee:1977tib,
Petcov:1976ff, Pal:1981rm, Shrock:1982sc, Dvornikov:2003js, Giunti:2014ixa}.

This mass suppression is avoided in extensions of the SM such as the one shown
in \cref{fig:feynman} (b). In this example, leptoquarks with $SU(3) \times
SU(2) \times U(1)$ quantum numbers of $(\mathbf{\bar{3}}, \mathbf{1}, 1/3)$ and
$(\mathbf{\bar{3}}, \mathbf{2}, 1/6)$ are introduced. The former one, called
$S_1$, could explain the $R(D^{(*)})$ and muon $g-2$ anomalies
(see refs.~\cite{Lees:2013uzd, Bauer:2015knc, Hirose:2016wfn, Aaij:2015yra, Aaij:2014ora,
  Aaij:2017vbb, Aaij:2013qta, Aaij:2015oid, Aaij:2019wad, Buttazzo:2017ixm,
Dorsner:2019itg, Brdar:2020quo} and \cref{sec:g-2}),
while the second one is introduced in the
context of the Voloshin mechanism \cite{Voloshin:1987qy} to avoid unacceptably
large contributions to the neutrino mass matrix without tuning. It is
intriguing that models of TeV-scale new physics that have been proposed for
other reasons can naturally lead to sizeable neutrino magnetic moments. For
instance, based on naive dimensional analysis, the diagram in
\cref{fig:feynman}~(b) predicts $\mu_\nu \sim e \, m_b / (16\pi^2 m_\text{LQ}^2)
\sim \SI{e-11}{\mu_B}$, where $m_\text{LQ} \sim \si{TeV}$ is the mass of the
leptoquark, $e$ is the electromagnetic gauge coupling, $m_b$ is the bottom quark mass
(responsible for the chirality flip in the diagrams), and $\mu_B = e / (2 m_e)$
is the Bohr magneton, the conventional unit for magnetic moments. As we will
see in the following sections, such values of $\mu_\nu$ are well within reach
of current experiments, depending on the mass of the $N_R$.

\section{Direct Detection of Neutrino Magnetic Moments}

\begin{figure}
  \centering
  \includegraphics[width=0.6\textwidth]{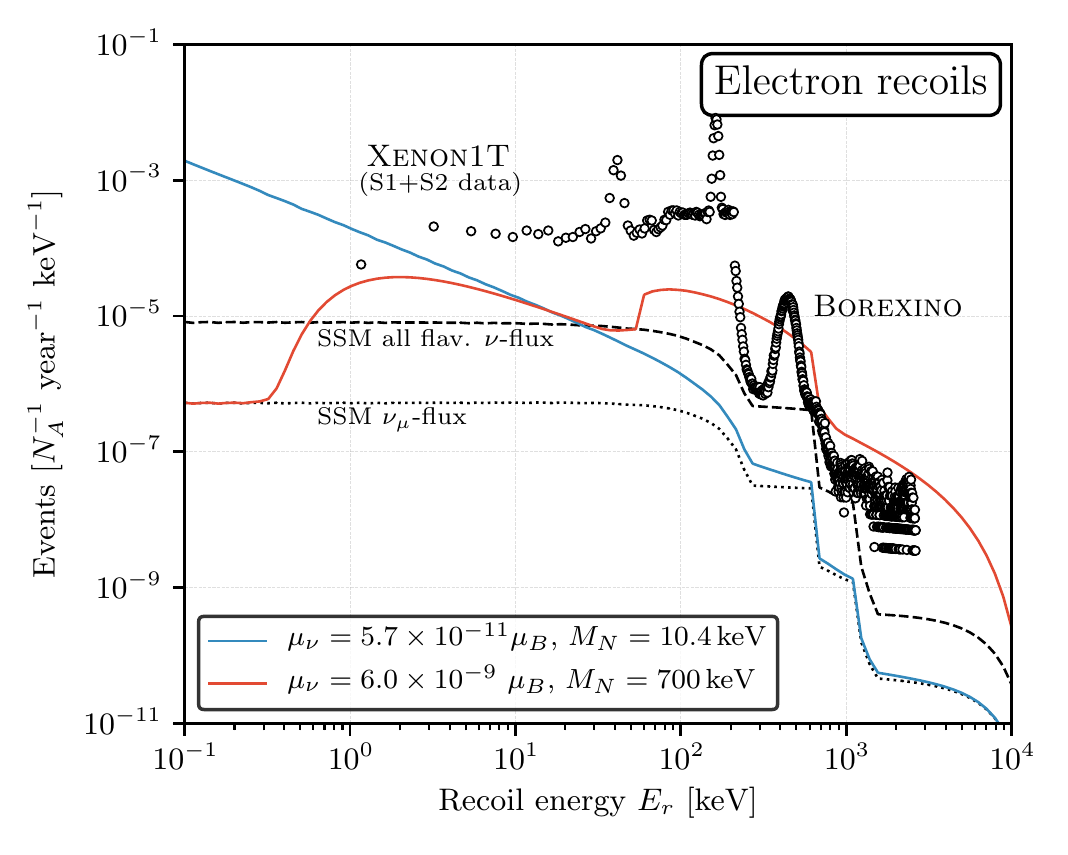}
  \caption{Spectrum of neutrino-induced electron recoils for the
    Standard Solar Model (SSM, dotted and dashed black lines) and in two
    different scenarios with non-negligible neutrino magnetic moments
    (blue and red lines). We also compare to the data from Borexino
    \cite{Agostini:2018uly} and Xenon1T \cite{Aprile:2020tmw}.}
  \label{fig:dd}
\end{figure}

A particularly promising way of detecting neutrino magnetic moments is offered
by detectors searching for dark matter scattering on nuclei and electrons.  As
is well known, solar neutrinos are an irreducible background to these searches
\cite{Gutlein:2010tq, Harnik:2012ni, Feng:2014uja}.
Interestingly, for non-negligible $\mu_\nu$, the solar neutrino scattering rate
is enhanced $\propto 1/E_r$ at low recoil energies $E_r$ due to the process
being mediated by a massless photon. This makes dark matter detectors with their
low recoil energy thresholds ideal tools to search for neutrino magnetic
moments.  This is illustrated in \cref{fig:dd}, where we compare the expected
electron recoil spectrum with and without neutrino magnetic moments to the
data from Xenon1T \cite{Aprile:2020tmw} and Borexino \cite{Agostini:2018uly}
(see also refs.~\cite{Harnik:2012ni, Babu:2020ivd, Shoemaker:2020kji}).
If the right-handed neutrino states are very light (blue line), we see a pronounced
enhancement of the event rate at low recoil energies, relevant to Xenon1T and
other dark matter detectors, while at higher energies, where dedicated solar
neutrino experiments like Borexino are sensitive, the magnetic moment
effect is negligible. For heavier right-handed neutrinos (red line), the
scattering kinematics are modified such that the kinks in the spectrum
(originating from the kinematic thresholds of the different components of
the solar neutrino flux) are shifted. In this case, significant
deviations from the Standard Model prediction are possible also at
larger energies.

\begin{figure}
  \centering
  \includegraphics[width=0.8\textwidth]{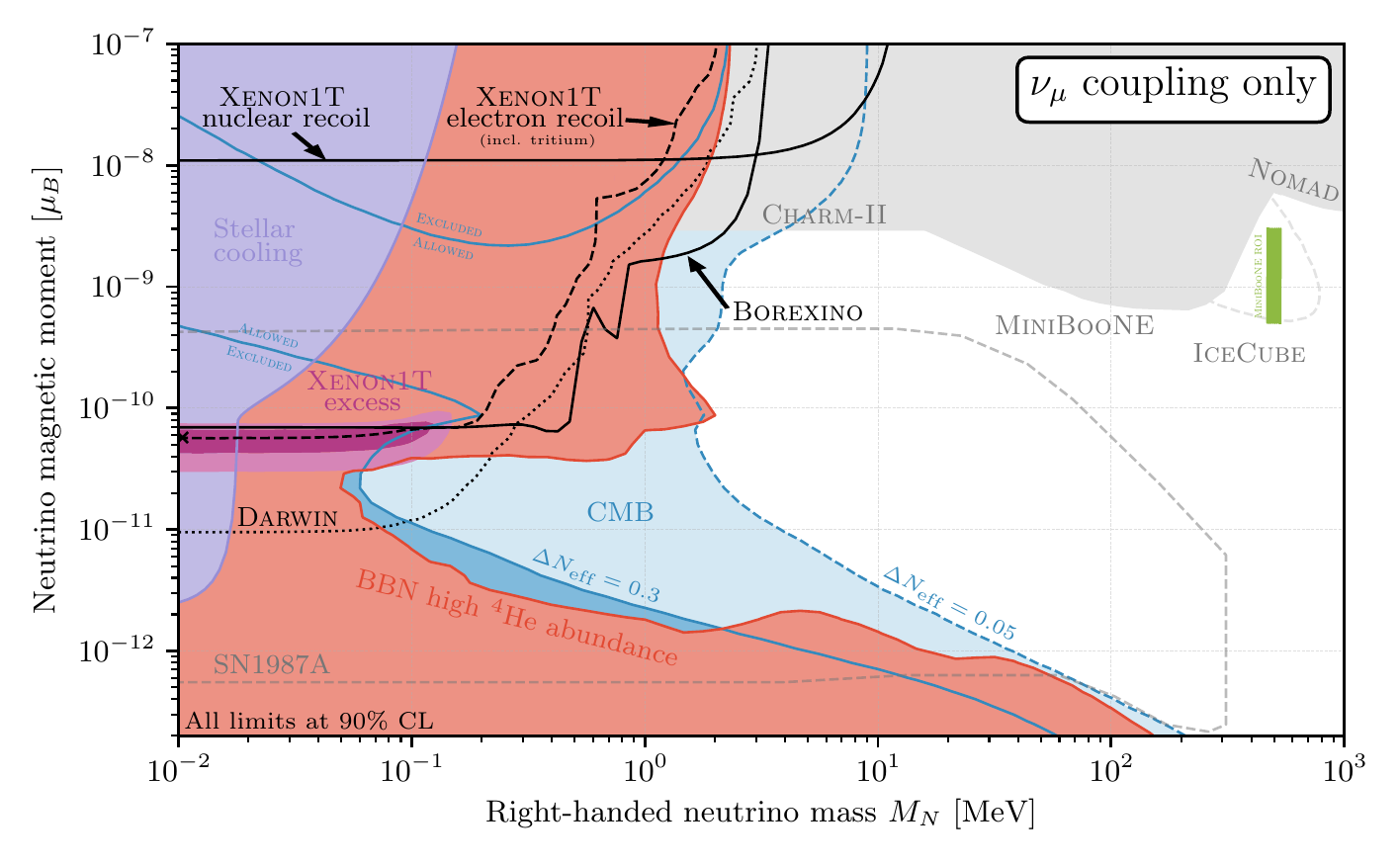}
  \caption{Summary of constraints on neutrino magnetic moments as a function
    of the right-handed neutrino mass. We show limits from Xenon1T (dashed line
    and upper solid black line) and from Borexino (lower solid black line), as
    well as the expected sensitivity of DARWIN (dotted black).  The Xenon1T
    electron recoil excess can be explained in the violet region at $\mu_\nu
    \sim \text{few} \times \SI{e-11}{\mu_B}$. Stellar cooling excludes the
    region shown in purple, while SN~1987A is sensitive to the region
    delineated by the dashed gray line.  Cosmology disfavors the red region via
    BBN, and the blue region through the $N_\text{eff}$ measurement in the CMB.
    Fixed-target accelerator experiments exclude the upper gray regions. We
    finally show the region favored by some explanation attempts for the
    MiniBooNE anomaly (green), and the sensitivity of IceCube via double-bang
    signatures~\cite{Coloma:2017ppo}. We comment in the text on the robustness
    of the various constraints.}
  \label{fig:constraints}
\end{figure}

Translating these results into constraints, we see in \cref{fig:constraints}
that Xenon1T is currently the most sensitive terrestrial probe of
neutrino magnetic moments for low right-handed neutrino mass $M_N$,
while at higher $M_N$, Borexino takes over.

\section{Astrophysical, Cosmological, and Accelerator Constraints}

While direct dark matter searches are the most sensitive \emph{terrestrial}
probes of neutrino magnetic moments, astrophysical and cosmological
constraints can be even stronger in large regions of parameter space.

\subsection{Stellar Cooling}

In the hot plasma inside a star, the photon dispersion relation is modified
in a way that can be interpreted as photons (or rather their in-medium
counterparts called plasmons) acquiring a mass on the order of
the plasma temperature $T$. Therefore, plasmon decays to $\nu_L + N_R$, mediated
by the magnetic moment coupling, become possible if $T \gtrsim M_N$.
Since neutrinos can leave the star unimpeded, this leads to significant
energy losses \cite{Raffelt:1996wa,
Raffelt:1999gv}. These losses modify stellar evolution and change, for instance,
the mass of the helium core of red giant stars at the time of the helium
flash.  The latter, in turn, can be observed through the luminosity at the
tip of the red giant branch in the Hertzsprung--Russel diagram
\cite{Raffelt:1994ry, Raffelt:1996wa,
Arceo-Diaz:2015pva, Diaz:2019kim}.
The resulting constraint, shown in purple in \cref{fig:constraints},
excludes neutrino magnetic moments down to $\mu_\nu \sim \text{few}
\times \SI{e-12}{\mu_B}$ for light $M_N$, but peters out at $M_N \gtrsim
\SI{10}{keV}$, the typical core temperature of a red giant star.
The kink in the exclusion region is due
to an additional energy loss process, $\gamma + e^- \to
\nu_L + N_R + e^-$, which extends the sensitivity to somewhat larger
$M_N$.

Note that stellar cooling constraints can be avoided in so-called
``chameleon'' models, in which $M_N$ depends dynamically on
the surrounding matter density.

\subsection{Supernova 1987A}

The processes that lead to anomalous cooling of stars are also active in
supernova cores, where they accelerate the cooling rate of the nascent neutron
star \cite{Raffelt:1996wa, Magill:2018jla}. The latter is constrained by the
observed duration of the neutrino burst from supernova 1987A, and the resulting
constraint is delineated by the gray dashed line in \cref{fig:constraints}.
Note the characteristic wedge shape of the excluded region, which is a
reflection of the fact that for too large $\mu_\nu$, neutrinos will no longer
be able to leave the supernova core without scattering and re-depositing their
energy.

We remark that constraints from SN~1987A have recently been called
into question by the authors of ref.~\cite{Bar:2019ifz}, who argue that
the observed neutrinos might not have originated from the cooling
neutron star itself, but from the accretion flow surrounding it. In
this case, they would carry no information on the cooling rate of the
neutron star, and the limit would disappear.

\subsection{Cosmology}

For not too small $\mu_\nu$, the neutrino magnetic moment operator leads
to the production of $N_R$ in the early Universe.  The presence of $N_R$ has
manifold consequences on the evolution of the Universe, especially during
BBN:
\begin{itemize}
  \item an increase in the energy density of relativistic species (measured
    through the parameter $N_\text{eff}$) if $M_N$ is below the BBN
    temperature, $T_\text{BBN} \sim \SI{1}{MeV}$.  The resulting increased
    expansion rate implies that $p \leftrightarrow n$ interactions freeze out
    faster and that neutrons have less time to decay.  Both effects imply a
    larger abundance of heavy elements.
  \item a phase of full or partial matter domination around BBN if $M_N \gg
    T_\text{BBN}$, but the $N_R$ lifetime is longer than a few minutes.
  \item a decrease in the baryon-to-photon ratio $\eta $.  If $N_R$ decays
    to photons and $\nu_L$ happen after BBN, this decrease in $\eta$
    must be compensated by a larger $\eta$ during BBN as $\eta$ is measured
    very precisely in the CMB, The larger $\eta$ during BBN renders
    deuterium disintegration less efficient and therefore production of
    heavy elements more efficient.
  \item a change in the photon-to-neutrino ratio at the CMB epoch
    (measured through $N_\text{eff}$) if the $N_R$ decay after neutrino
    decoupling.
\end{itemize}
Using a modified version of the AlterBBN code \cite{Arbey:2011nf,
Arbey:2018zfh, Depta:2020wmr}, we have simulated BBN in the presence
of right-handed neutrinos and non-zero $\mu_\nu$.  Comparing the
predicted primordial abundances of light elements, especially \iso{He}{4},
with observations, we obtain the exclusion region shown in red in
\cref{fig:constraints}.  At $M_N \lesssim \si{MeV}$, or at low $\mu_\nu$
and thus long $N_R$ lifetime, the $N_R$ are present during BBN, increasing
the expansion rate of the Universe. Moreover, their eventual decay injects
extra energy into the plasma after BBN. Note that there is a sweet spot
at masses $\lesssim \si{MeV}$ and $\mu_\nu \sim \SI{e-11}{\mu_B}$, where
the limit is relatively weak. There, $N_R$ decouple around the
QCD phase transition so that their abundance is significantly diluted
by entropy production during the QCD phase transition, and at the same time their mass
is sufficiently small to avoid a phase of full or partial matter domination.

However, when using the same simulation that predicts the \iso{He}{4}
abundance to also predict the value of $N_\text{eff}$ at recombination,
we see that this sweet spot is closed.  For instance, the parameter region
favored by the Xenon1T excess is completely closed if the current limit
$\Delta N_\text{eff} < 0.05$ \cite{Philcox:2020vvt} is taken at face value.
If we allow for larger $\Delta N_\text{eff} \simeq 0.2$, as suggested
by the tension between primordial and late-time measurements of the
Hubble constant, the Xenon1T region may still be marginally allowed.

Let us finally remark that cosmological limits are quite robust. Avoiding
or relaxing them requires either a mechanism that prevents $N_R$ production
in the early Universe altogether, for instance by coupling $N_R$ to
a slowly rolling scalar field, such that the $N_R$ mass in the early
Universe is different from the one today.  Or one could mitigate the
problems created by $N_R$ decay to $\nu_L + \gamma$
by introducing a second decay mode to lighter SM singlets.

\subsection{Accelerator Constraints}

Coming back to Earth, an additional set of constraints on neutrino magnetic
moments can be obtained from direct searches for $N_R$ production in fixed-target
accelerator experiments. \Cref{fig:constraints} shows in gray the
limits derived from Charm-II, MiniBooNE, and NOMAD \cite{Shoemaker:2018vii}.
In the case of MiniBooNE, we also indicate in green the parameter region favored
by certain attempts to explain the MiniBooNE anomaly \cite{Gninenko:2009ks,
Magill:2018jla}.

\section{The Muon $g-2$}
\label{sec:g-2}

We finally comment on the relation between neutrino magnetic moments and the
anomalous magnetic moment of the muon, $(g-2)_\mu$. We focus on the leptoquark scenario
we have presented as an example in~\cref{sec:mm}.  In the Lagrangian of the
$S_1$ leptoquark, the terms relevant to the generation of $\mu_\nu$
through the diagram in \cref{fig:feynman}~(b) are
\begin{align}
  \mathcal{L}_{S_1}
    \supset y_1 \, \overline{b_R^{\ c}} N_R \, S_1
          + y_2 \, \overline{Q_{L}^3} L^{i\; c}_L \, S^\dagger_1 + \text{h.c.}\,.
  \label{eq:L-lq-1}
\end{align}
The second of these also generates loop contributions to $(g-2)_\mu$, but naive
dimensional analysis shows that these are smaller by about an order of magnitude
than the observed discrepancy between theoretical predictions~\cite{Aoyama:2020ynm}
and the BNL/Fermilab results~\cite{Bennett:2006fi,Abi:2021gix}. However,
the $S_1$ leptoquark contributes to the muon's $g-2$ also through another
interaction, namely the operator $\mathcal{L} \supset y'_1\, \overline{t_R^{\ c}}
e^i_R \, S_1$.  This operator generates loops involving top quarks, and because
having a top quark in the loop lifts the chiral suppression that normally
plagues loop contributions to the muon $g-2$, even a tiny coupling
$y'_1$ is sufficient to accommodate the experimental results.

\section*{Acknowledgments}

This work has been supported by the European Research Council (ERC) under the
European Union's Horizon 2020 research and innovation program (grant agreement
No.\ 637506, ``$\nu$Directions'' and grant agreement No.\ 833280, ``FLAY'').
Fermilab is operated by the Fermi Research Alliance, LLC under contract
No.~DE-AC02-07CH11359 with the US~DOE.

\bibliographystyle{kpmoriond}
\bibliography{refs}

\end{document}